\documentclass[10pt,conference]{IEEEtran}
\usepackage[caption=false,font=normalsize,labelfont=sf,textfont=sf]{subfig}
\usepackage{graphicx}
\usepackage{stfloats} 
\IEEEoverridecommandlockouts
\usepackage{hyperref}
\hypersetup{
    colorlinks=true,
    linkcolor=black,
    citecolor=black,
    urlcolor=black
}
\usepackage{cite}
\usepackage{amsmath,amssymb,amsfonts}
\usepackage{algorithmic}
\usepackage{graphicx}
\usepackage{textcomp}
\usepackage{xcolor}

\def\BibTeX{{\rm B\kern-.05em{\sc i\kern-.025em b}\kern-.08em
    T\kern-.1667em\lower.7ex\hbox{E}\kern-.125emX}}
\begin{document}

\title{Doppler-Based Multistatic Drone Tracking via Cellular Downlink Signals \vspace{-0.5cm}
\thanks{
This work was supported in part by the National Natural Science Foundation of China (NSFC) under Grant 62171213 and Grant 62522107, in part by Shenzhen Science and Technology Program under Grant JCYJ20241202125328038, and in part by High Level of Special Funds under Grant G03034K004. \textit{(Corresponding author: Rui Wang)}

}
}


\author
{\IEEEauthorblockN{Chenqing Ji\IEEEauthorrefmark{1}, Qionghui Liu\IEEEauthorrefmark{1}, Jiahong Liu\IEEEauthorrefmark{1}, Chao Yu\IEEEauthorrefmark{1}, Yifei Sun\IEEEauthorrefmark{1}, Rui Wang\IEEEauthorrefmark{1} and Fan Liu\IEEEauthorrefmark{2}}
\IEEEauthorblockA{\IEEEauthorrefmark{1}Department of Electronic and Electrical Engineering, College of Engineering, Southern University \\
of Science and Technology, Shenzhen 518055, China \\
\IEEEauthorrefmark{2}School of Information Science and Engineering, Southeast University, Nanjing 210096, China \\
Email: \{12332152,12212010,12210631,12431241,sunyf2019\}@mail.sustech.edu.cn, \href{wang.r@sustech.edu.cn}{wang.r@sustech.edu.cn}, \href{fan.liu@seu.edu.cn}{fan.liu@seu.edu.cn}
}
}

\maketitle

\begin{abstract}
In this paper, a multistatic Doppler sensing system is proposed for the drone tracking via downlink Long-Term Evolution (LTE) signals. Specifically, the LTE base stations (BSs) are exploited as signal illuminators, and three passive sensing receivers are deployed at different locations to detect the bistatic Doppler frequencies of a target drone from received downlink signals. It is shown that even without the measurements of BS-drone-receiver range and angle, the Doppler measurements could provide sufficient information for trajectory tracking. Particularly, the trajectory of the target drone, consisting of the initial position and velocities of all the time slots, can be reconstructed by solving a minimum mean-squared error problem according to the above Doppler measurements. It is demonstrated by experiment that although the target drone and all the sensing receivers are around $\mathbf{200}$ meters away from the illuminating BSs, the complicated trajectories can be tracked with $\mathbf{90\%}$ errors below $\mathbf{90}$ centimeters. Since this accuracy is notably higher than the typical range resolution of LTE signals, the demonstration shows that drone trajectory tracking with a high accuracy could be feasible solely according to Doppler detection, as long as the deployment density of receivers is sufficiently high.
\end{abstract}

\begin{IEEEkeywords}
Integrated sensing and communication, multistatic sensing, drone tracking.
\end{IEEEkeywords}

\section{Introduction}
With the development of the low-altitude economy, various types of drones have been widely adopted in many applications, which have spurred extensive research efforts to the area of drone monitoring, especially the small drones at a low altitude. Recently, integrated sensing and communication (ISAC) technology, which utilizes the cellular communication signals in environment sensing, has been becoming a promising solution for the surveillance of low-altitude drones. 

In fact, numerous studies have been devoted to the passive sensing technology for drone detection. They leveraged various broadcast signals, such as digital audio broadcasting (DAB) \cite{7944357}, digital video broadcasting—terrestrial (DVB-T) \cite{8546549,8768179} and digital video broadcasting—satellite (DVB-S) \cite{9266624,9114795}, as illumination sources. These works demonstrated the feasibility of passive sensing technology in drone monitoring. However, compared with traditional TV or radio towers, cellular base stations (BSs) are more densely deployed, enabling fine-grained drone detection. Therefore, there are also a number of studies exploiting various cellular signals in small drone detection, including the global system for mobile communications (GSM) signals, the Long-Term Evolution (LTE) signals, the fifth-generation (5G) mobile communication signals, etc. For instance, in \cite{7497375}, a GSM-based passive coherent location (PCL) system was proposed to detect drones with low radar cross-section (RCS) by analyzing the Doppler frequency from the scattered GSM signals. In \cite{10902137}, a passive radar system was developed to detect a drone in complex electromagnetic environments by leveraging the LTE signals from three non-cooperative transmitters. Furthermore, an entropy-based adaptive integration method using the 5G signals in passive radar was proposed in \cite{rs14236146}, which could enhance the drone detection accuracy in dynamic environments by selecting the optimal time segments. All the above works focused on the existence detection of drones, instead of trajectory tracking.

The angle-of-arrival (AoA), bistatic range, and bistatic Doppler frequency at single or multiple passive sensing receivers can be jointly exploited in drone tracking \cite{9764210,10950409}. However, a large antenna aperture is usually required to achieve a high angular resolution, and minor deviation in the angle measurements can lead to considerable localization errors at a long distance \cite{8692423}. Moreover, since the range resolution depends on signal bandwidth, higher resolution requires larger downlink bandwidth \cite{griffiths2022introduction}. For example, LTE signals with a $20$MHz bandwidth provide a resolution of $7.5$ meters, which is far below the requirements for high-accuracy tracking. Fortunately, it is much easier for the sensing receiver to obtain a high-accuracy measurement of the Doppler frequency, which is independent of antenna aperture and signal bandwidth. Hence, it is interesting to ask if it is possible to achieve high-accuracy drone tracking via the sensing of Doppler frequency. 

In this paper, a multistatic sensing system is proposed to track the trajectory of a target drone with LTE downlink signals. In order to avoid the high hardware cost and low resolution measurements of AoA and range, a novel optimization method is proposed to reconstruct the drone's trajectory via the Doppler measurements of three sensing receivers. Particularly, the proposed system consists of three passive sensing receivers at distinct locations, each equipped with two radio frequency chains to detect the bistatic Doppler frequency of the target drone. Based on the Doppler measurements, the estimation of drone trajectory is formulated as a minimum mean-squared error problem, and a gradient-descent method is adopted to reconstruct the trajectory. Experimental results demonstrate that without the knowledge on the initial location of the drone, $90\%$ of the tracking errors are below $0.9$ meters, when the receiver-drone distances are around $20$ meters and BS-drone distances are around $200$ meters. 

The remainder of this paper is organized as follows. An overview of the proposed drone tracking system is provided in Section \ref{sec2}.  The signal model and the detection technique for the bistatic Doppler frequency are presented in Section \ref{sec3}.  The motion model and drone tracking method are elaborated in Section \ref{sec4}.  The experimental platform and the tracking results are demonstrated in Section \ref{sec5}. Finally, the conclusion is drawn in Section \ref{sec6}.

\section{System Overview} \label{sec2}
\begin{figure}[!ht]
    \centering
    \vspace{-0.5cm}
    \includegraphics[width=0.48\textwidth
    ]{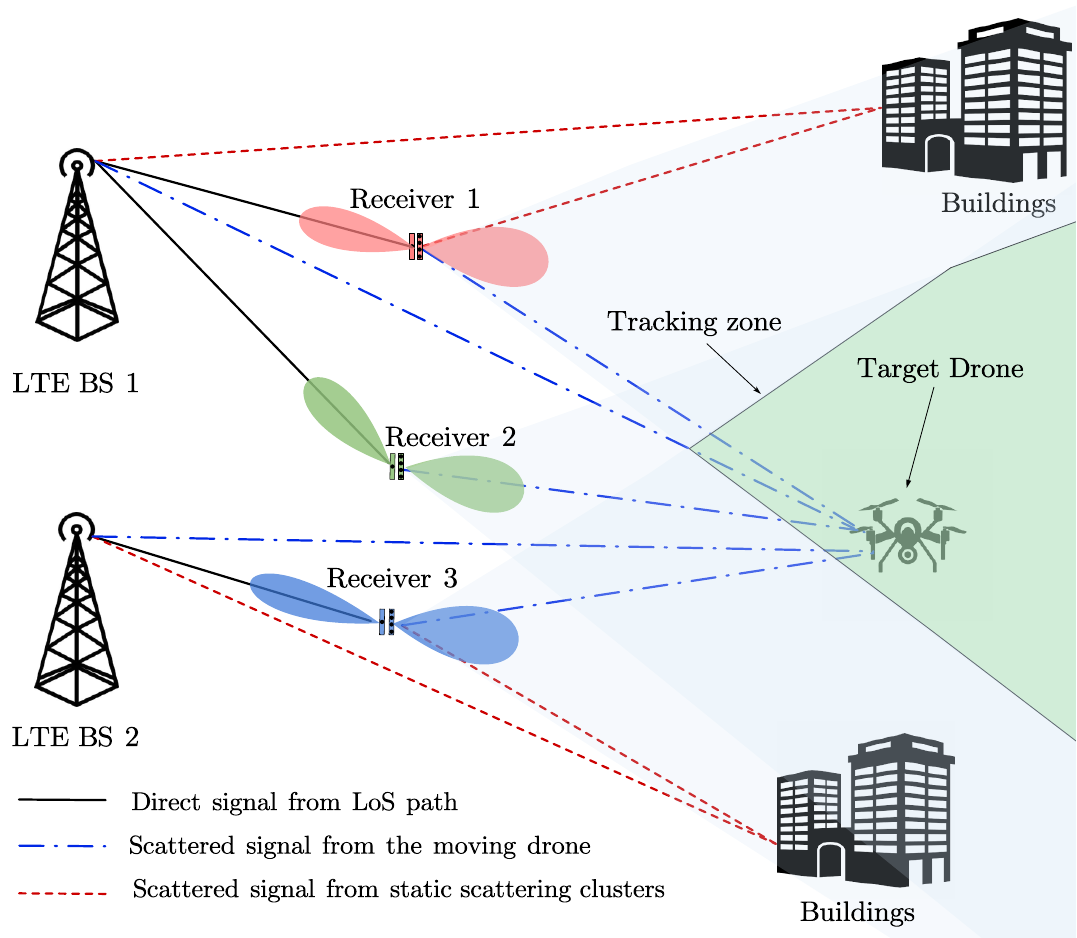}
     \vspace{-5mm}
    \caption{An example scenario of passive drone tracking.}
    \label{fig1}
\end{figure}

The proposed passive drone tracking system consists of one or multiple LTE BSs and at least three receivers. It is deployed to track the trajectory of a target drone flying on a two-dimensional plane at a fixed altitude. 
Without loss of generality, the scenario with two LTE BSs, as illustrated in Fig. \ref{fig1}, is elaborated in this paper, which is the same as the experiment scenario in Section \ref{sec5}. 
The two BSs and the three receivers are referred to as the BS 1, BS 2, Receiver 1, Receiver 2 and Receiver 3, respectively. Due to the frequency reuse mechanism, the two BSs work on two different frequency bands.
The Receiver 1 and 2 receive the signals of the BS 1, and the Receiver 3 receives the signals of the BS 2, respectively. 

The three receivers are deployed to detect the bistatic Doppler frequencies of one target drone. Hence, there are two radio frequency (RF) chains at each receiver. One narrow receive beam is directed to its corresponding BS, and the other wide receive beam is targeted at the drone tracking zone, as shown in Fig. \ref{fig1}. For the elaboration convenience, the line-of-sight (LoS) signal path between the BS $i$ ($i=1,2$) and its corresponding Receiver $j$ ($j=1,2,3$) is referred to as the ($i,j$)-th reference channel, while the corresponding BS-drone-receiver signal path is referred to as the ($i,j$)-th surveillance channel. By correlating the received signals of both channels, each receiver can detect the bistatic Doppler frequency of the target drone at one direction. Since the bistatic Doppler frequency detection depends on time-frequency characteristics of the opportunistically received signals rather than demodulating the information-bearing contents, Doppler detection remains feasible even without active user connections. Relying only on the detected Doppler frequencies at the three receivers, the trajectory of the target drone can be reconstructed, even when the initial location of the drone is unknown. 

Although the downlink signals of two LTE BSs are exploited in the considered scenario, the proposed method can be used in the scenario with single or three BSs, as long as both the reference and surveillance channels of one receiver originate from the same BS. Moreover, although the drone's motion is constrained to a fixed altitude, the proposed method can be extended to three-dimensional tracking with more receivers.

\section{Bistatic Doppler frequency detection} \label{sec3}
\subsection{Signal Model}
As illustrated in Fig. \ref{fig1}, there are three pairs of reference channel and surveillance channel in the system. Without loss of generality, the bistatic Doppler frequency detection between the BS $i$ and the Receiver $j$ is elaborated in this section\footnote{When $i=1$, $j=1,2$; when $i=2$, $j=3$ in Fig. \ref{fig1}.}. Let $s_{i}(t)$, $ t\in [0, \mathrm{T}]$, be the information-bearing signal transmitted from the BS $i$ ($i=1,2$). The received signal of the $(i,j)$-th reference channel, denoted as $y_{r,j}(t)$, can be expressed as
\begin{equation}
y_{r,j}(t)=\alpha_{r,j}s_{i}\left(t-\tau_{r,j}\right)+n_{r,j}(t),
\label{eq1}
\end{equation}
where $\alpha_{r,j}$ and $\tau_{r,j}$ denote the complex gain and delay of the LoS path respectively, and $n_{r,j}(t)$ denotes the noise and interference (the downlink signals scattered off the surrounding buildings). 
Due to the loss of scattering, the power of $n_{r,j}(t)$ in (\ref{eq1}) is usually much weaker than that of LoS signal.

Meanwhile, the received signal of the $(i,j)$-th surveillance channel, denoted as $y_{s,j}(t)$, includes the scattered signals off the target drone and the surrounding static buildings. It can be expressed as
\begin{equation}
\begin{aligned}
y_{s,j}(t) & =\alpha_{s,j}^{\mathrm{Tar}}(t)s_{i}\left(t-\tau_{s,j}^{\mathrm{Tar}}(t)\right)e^{j2\pi f_{j}^{\mathrm{Tar}}(t)t} \\
 & +\sum_{l=1}^{L_{s,j}}\alpha_{s,j}^ls_{i}\left(t-\tau_{s,j}^l\right)+n_{s,j}(t), \label{eq2}
 \end{aligned}
\end{equation}
where $\alpha_{s,j}^{\mathrm{Tar}}$, $\tau_{s,j}^{\mathrm{Tar}}(t)$ and $f_{j}^{\mathrm{Tar}}(t)$ denote the 
complex gain, delay and Doppler frequency of the scattered signals off the target drone respectively, $L_{s,j}$ denotes the number of undesired paths, $\alpha_{s,j}^l$ and $\tau_{s,j}^l$ denote the complex gain and delay of the $l$-th one, and $n_{s,j}(t)$ denotes the noise. Note that the signal from LoS path may also be received in the
surveillance channel, which is considered in the second term of (\ref{eq2}).

The received signals from both the reference channel and the surveillance channel at the three receivers are sampled with a period $\mathrm{T_s}$, which can be written as $$y_{r,j}[n]=y_{r,j}(n\mathrm{T}_\mathrm{s})$$ and $$y_{s,j}[n]=y_{s,j}(n\mathrm{T}_\mathrm{s}),$$ respectively, where $n = 1, ..., \mathrm{T}/\mathrm{T_s}$ is the sample index.
As a remark, the signal components with zero Doppler frequency in $y_{s,j}[n]$ 
may interfere with the detection of the target drone's Doppler frequency. Therefore, the least-square-based (LS-based) adaptive interference cancellation in \cite{doi:10.1049/ip-rsn:20055038} is utilized to suppress the above interference. The signal of the surveillance channel at the Receiver $j$ after interference cancellation is denoted as $\hat{y}_{s,j}[n]$ ($j=1,2,3$).

\subsection{CAF-Based Detection}

The detection of time-varying bistatic Doppler frequency of the target drone is based on the cross-ambiguity function (CAF) between the received signals of the reference channel and the surveillance channel.
Specifically, the bistatic Doppler frequency of the target drone at each receiver is detected every $\mathrm{T_d}$ seconds, and the time instances $t=k\mathrm{T_d}$ ($k=1,2,...,K$) are referred to as the detection time instances. At each detection time instance, a window of $N_{w}$ samples is utilized to calculate the CAF. As a remark, the Doppler frequency is assumed to remain approximately constant within this window. Therefore, the CAF at the Receiver $j$ and the $k$-th detection time instance can be calculated as
\begin{equation}
R_{j,k}(f_D)=\max_{\tau_{j,k}}\sum_{n=(k-1)N_0+1}^{(k-1)N_0+N_w}\hat{y}_{s,j}[n]y_{r,j}^*[
n-\tau_{j,k}]e^{-j2\pi f_Dn\mathrm{T}_\mathrm{s}}, \label{eq3}
\end{equation}
where $N_{0} = \mathrm{T_{d}/T_{s}}$, and $(.)^*$ is the complex conjugate. Due to the limited bandwidth of cellular signals, the propagation delay $\tau_{j,k}$, can be neglected\footnote{As shown in our experiment that the tracking error is below $0.9$m, which is much smaller than the range resolution of the typical LTE signal bandwidth $20$MHz. Therefore, the propagation delay $\tau_{j,k}$ is not exploited in this work.}. Moreover, it can be observed that a peak value of $R_{j,k}(f_D)$ can be detected at $f_D = f_{j}^\mathrm{Tar}{(k\mathrm{T}_\mathrm{d})}$. In other words, the bistatic Doppler frequency of the drone can be detected by finding the peak value of $R_{j,k}(f_D)$.

In practice, there might be more than one peak values of $R_{j,k}(f_D)$ in a detection time instance due to noise and interference, namely false alarm. To suppress the false alarms, an adaptive threshold-based filtering method is applied. Particularly, a bistatic Doppler frequency $f_{D}$ is detected at the Receiver $j$ and the $k$-th detection time instance if
$$\left|R_{j,k}(f_D)\right|\geq\beta_{j,k}(f_D).$$ The detection threshold $\beta_{j,k}(f_D)$ is determined according to \cite{10379434} as
\begin{equation}
\beta_{j,k}(f_{D})=\frac{\gamma}{2C+1}\sum_{c=-C}^{C}\big|R_{j,k}(f_{D}+c\Delta f)\big|,
\end{equation}
where $\gamma > 1$ is a
scaling factor, $C$ is the half length of training cells, $\Delta f=\frac{1}{N_w\mathrm{T_s}}$ is the resolution of the Doppler frequency.

With the above detection method, one estimation of bistatic Doppler frequency per detection time instance can be determined based on the maximum amplitude at each receiver, while the issue of the miss detection is fully mitigated via linear interpolation. Denote the estimated bistatic Doppler frequencies at the Receiver $j$ and the $k$-th detection time instance as $\hat{f}_{j,k}$, $j=1,2,3$, $k=1,2,...,K$. Finally, the Kalman filter is applied to smooth these estimations, and the smoothed estimations of bistatic Doppler frequency at the Receiver $j$ ($j=1,2,3$) are denoted as $\{\tilde{f}_{j,k}|\forall k\}$. 

\vspace{-0.1cm}
\section{Drone Trajectory Tracking} \label{sec4}
In this section, the method of drone trajectory tracking is elaborated according to the smoothed bistatic Doppler frequency estimations $\{\tilde{f}_{j,k}| \forall j,k\}$ at the three receivers. In the
following parts, we first establish the geometric relation between the bistatic Doppler frequencies at the three receivers and the motion parameters of the target drone, and then elaborate the drone tracking method.

\vspace{-0.1cm}
\subsection{Motion Model}
The location of the two LTE BSs and
the three receivers are denoted as $\mathbf{p}_{1}^{tx} = [x_{\mathrm{T1}}, y_{\mathrm{T1}}]^{\mathrm{T}}$, $\mathbf{p}_{2}^{tx} = [x_{\mathrm{T2}}, y_{\mathrm{T2}}]^{\mathrm{T}}$, $\mathbf{p}_{1}^{rx} = [0, 0]^{\mathrm{T}}$, $\mathbf{p}_{2}^{rx} = [x_{\mathrm{R2}}, y_{\mathrm{R2}}]^{\mathrm{T}}$, and $\mathbf{p}_{3}^{rx} = [x_{\mathrm{R3}}, y_{\mathrm{R3}}]^{\mathrm{T}}$,
respectively. They can be measured in advance with a high accuracy.
Let $\mathbf{p}_{1} = [x_{1}, y_{1}]^{\mathrm{T}}$ be the initial location of the target drone, $\mathbf{p}_{k} = [x_{k}, y_{k}]^{\mathrm{T}}$ and $\mathbf{v}_{k} = [v_{xk}, v_{yk}]^{\mathrm{T}}$ be the location and velocity of the target drone at the $k$-th detection time instance\footnote{For clarity, the height differences between the drone, the two BSs, and the three receivers can be neglected compared with their horizontal distances. 
}, the location update of the drone at the $(k+1)$-th detection time instance can be expressed as 
\begin{equation}
\mathbf{p}_{k+1} = \mathbf{p}_{k} + \mathbf{v}_{k} \mathrm{T_d}, k=1,2,...,K-1.
\end{equation}

Denote the aggregation of all the motion parameters up to the $k$-th detection time instance as $$\mathbf{m}_{k} = [\mathbf{p}_{1}, \mathbf{v}_{1}, \mathbf{v}_{2}, \mathbf{v}_{3}, ..., \mathbf{v}_{k}]^{\mathrm{T}}, k=1,2,...,K.$$
Let $f_{1,k}$, $f_{2,k}$ and $f_{3,k}$ be the actual bistatic Doppler frequencies of the drone at the $k$-th detection time instance and the Receiver 1, 2 and 3, respectively. We have
\begin{equation}
\begin{bmatrix}
f_{1,k}(\mathbf{m}_{k}) \\
f_{2,k}(\mathbf{m}_{k}) \\
f_{3,k}(\mathbf{m}_{k})
\end{bmatrix}=\mathbf{D}_k\cdot \mathbf{v}_k
, 
\end{equation}
where 
\begin{equation}
\mathbf{D}_k = -
\begin{bmatrix}
    \frac{1}{\lambda_1} \left( \frac{\mathbf{p}_{k} - \mathbf{p}_{1}^{tx}}{\|\mathbf{p}_{k} - \mathbf{p}_{1}^{tx}\|} + \frac{\mathbf{p}_{k} - \mathbf{p}_{1}^{rx}}{\|\mathbf{p}_{k} - \mathbf{p}_{1}^{rx}\|} \right)^{\mathrm{T}} \\
    \frac{1}{\lambda_1} \left( \frac{\mathbf{p}_{k} - \mathbf{p}_{1}^{tx}}{\|\mathbf{p}_{k} - \mathbf{p}_{1}^{tx}\|} + \frac{\mathbf{p}_{k} - \mathbf{p}_{2}^{rx}}{\|\mathbf{p}_{k} - \mathbf{p}_{2}^{rx}\|} \right)^{\mathrm{T}} \\
    \frac{1}{\lambda_2} \left( \frac{\mathbf{p}_{k} - \mathbf{p}_{2}^{tx}}{\|\mathbf{p}_{k} - \mathbf{p}_{2}^{tx}\|} + \frac{\mathbf{p}_{k} - \mathbf{p}_{3}^{rx}}{\|\mathbf{p}_{k} - \mathbf{p}_{3}^{rx}\|} \right)^{\mathrm{T}}
    \end{bmatrix}, \label{eq9}
\end{equation}
$\lambda_1$ and $\lambda_2$ denote the wavelength of the downlink signals of the BS 1 and 2 respectively. 

\subsection{Trajectory Tracking}
Let $\tilde{\mathbf{Z}}_{k} = [\tilde{\mathbf{z}}_1, \tilde{\mathbf{z}}_2, \tilde{\mathbf{z}}_3, ..., \tilde{\mathbf{z}}_k]^{\mathrm{T}}$ be the aggregation of the smoothed Doppler frequency estimations, where $\tilde{\mathbf{z}}_k = [\tilde{f}_{1,k}, \tilde{f}_{2,k}, \tilde{f}_{3,k}]^{\mathrm{T}}$ is the smoothed Doppler frequency estimations at the three receivers and the $k$-th detection time instance. Let $\mathbf{Z}_{k} = [\mathbf{z}_1, \mathbf{z}_2, \mathbf{z}_3, ..., \mathbf{z}_k]^{\mathrm{T}}$ be the aggregation of the actual Doppler frequencies, where $\mathbf{z}_{k} =[f_{1,k}(\mathbf{m}_{k}), f_{2,k}(\mathbf{m}_{k}), f_{3,k}(\mathbf{m}_{k})]^{\mathrm{T}}$, the mean-squared error of the Doppler frequency measurements is given by $$\varepsilon(\mathbf{m}_{K}) = \mathbf{tr}[(\mathbf{Z}_{K}-\tilde{\mathbf{Z}}_{K})^\mathrm{T}(\mathbf{Z}_{K}-\tilde{\mathbf{Z}}_{K})],$$
where $\mathbf{tr}[.]$ denotes the matrix trace. Hence, the estimation of the motion parameters of all the detection time instances can be formulated as 
\begin{equation}
\mathbf{m}_{K}^* =\underset{{\mathbf{m}}_{K}}{\operatorname*{\operatorname*{argmin}}} \: \varepsilon(\mathbf{m}_{K}). \label{eq8}
\end{equation}

The Levenberg-Marquardt (LM) algorithm in \cite{yu2024trajectorytrackingmmwavecommunication,7060406} can be applied to solve the above problem. Specifically, let $\tilde{\mathbf{m}}_{K,1}^0, \tilde{\mathbf{m}}_{K,2}^0, ..., \tilde{\mathbf{m}}_{K,S}^0$ be the $S$ initial solutions of $\mathbf{m}_{K}$. In the $p$-th iteration ($p = 1,2,3, ...$), the estimations of $\mathbf{m}_{K}$ can be updated as
\begin{equation}
     \tilde{\mathbf{m}}_{K,s}^p = \tilde{\mathbf{m}}_{K,s}^{p-1} - J(\tilde{\mathbf{m}}_{K,s}^{p-1})\Delta, s=1,2,3,...,S,
\end{equation}
where $J(\mathbf{m}_{K}) = \frac{\partial \varepsilon(\mathbf{m}_{K})}{\partial \mathbf{m}_{K}}$ is the Jacobian matrix and $\Delta$ is the step size. Let $P$ be the total number of iterations, and the final estimated motion
parameters of the drone can be expressed by
\begin{equation}
\mathbf{m}_{K}^* = \underset{{\mathbf{m}}_{K} \in \{\tilde{\mathbf{m}}_{K,1}^P, \tilde{\mathbf{m}}_{K,2}^P, ..., \tilde{\mathbf{m}}_{K,S}^P\}}{\operatorname{argmin}} \varepsilon({\mathbf{m}}_{K}).
\end{equation}
Finally, the trajectory of the drone can be reconstructed iteratively via $\mathbf{m}_{K}^* = [\mathbf{p}_{1}^*, \mathbf{v}_{1}^*, \mathbf{v}_{2}^*, \mathbf{v}_{3}^*, ..., \mathbf{v}_{K}^*]^{\mathrm{T}}$ as follows:
\begin{equation}
\mathbf{p}_{k+1}^* = \mathbf{p}_{k}^* + \mathbf{v}_{k}^* \mathrm{T_d}, k=1,2,...,K-1.
\end{equation}


\begin{figure}[!h]
    \centering
     \hspace{0.05cm}
    \subfloat[The stadium layout in the experiment]{
        \includegraphics[width=0.438\textwidth]{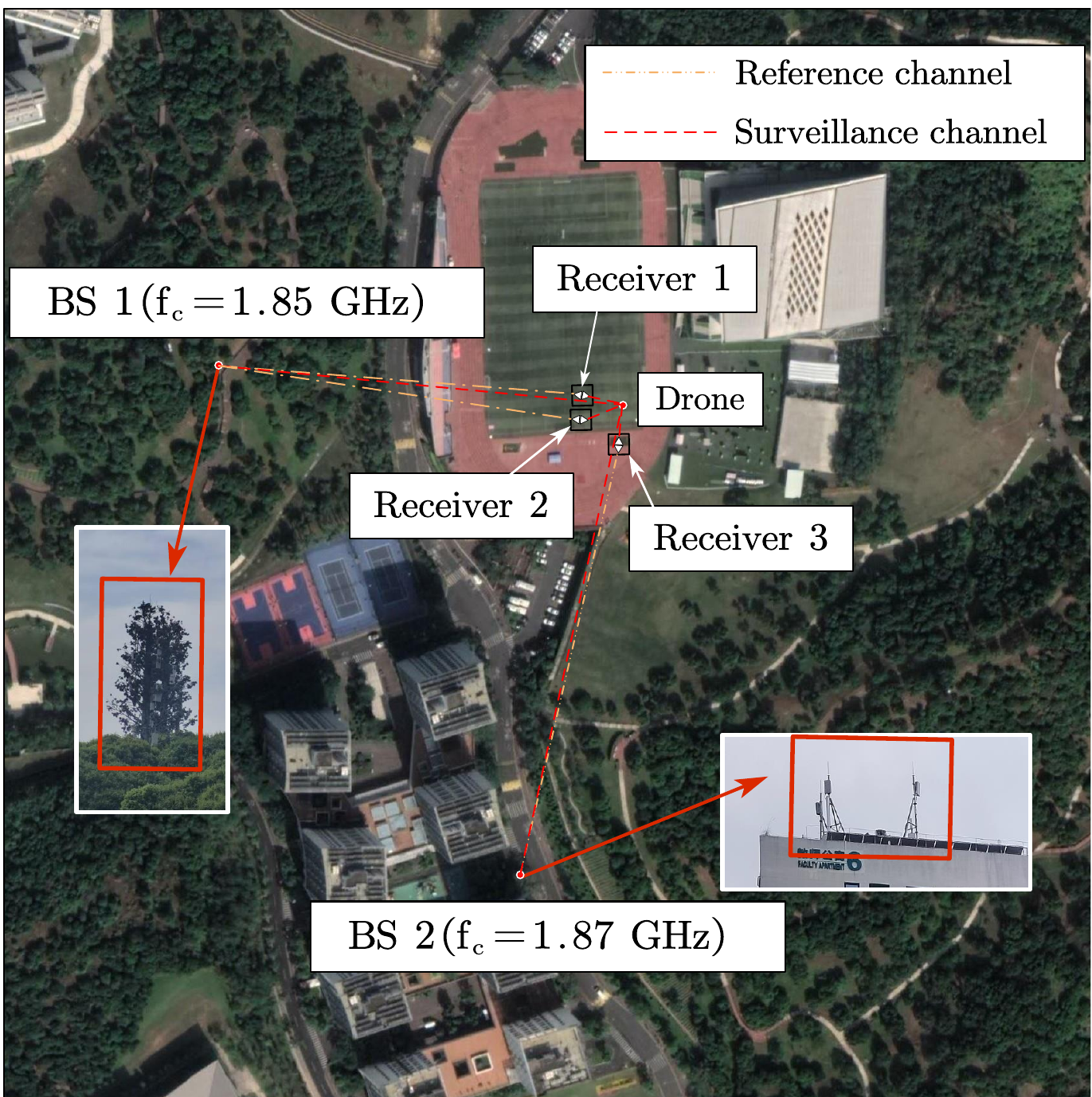}
    \label{fig4-a}}
     \hfill
    \subfloat[An snapshot of experiment scenario]{
        \includegraphics[width=0.44\textwidth]{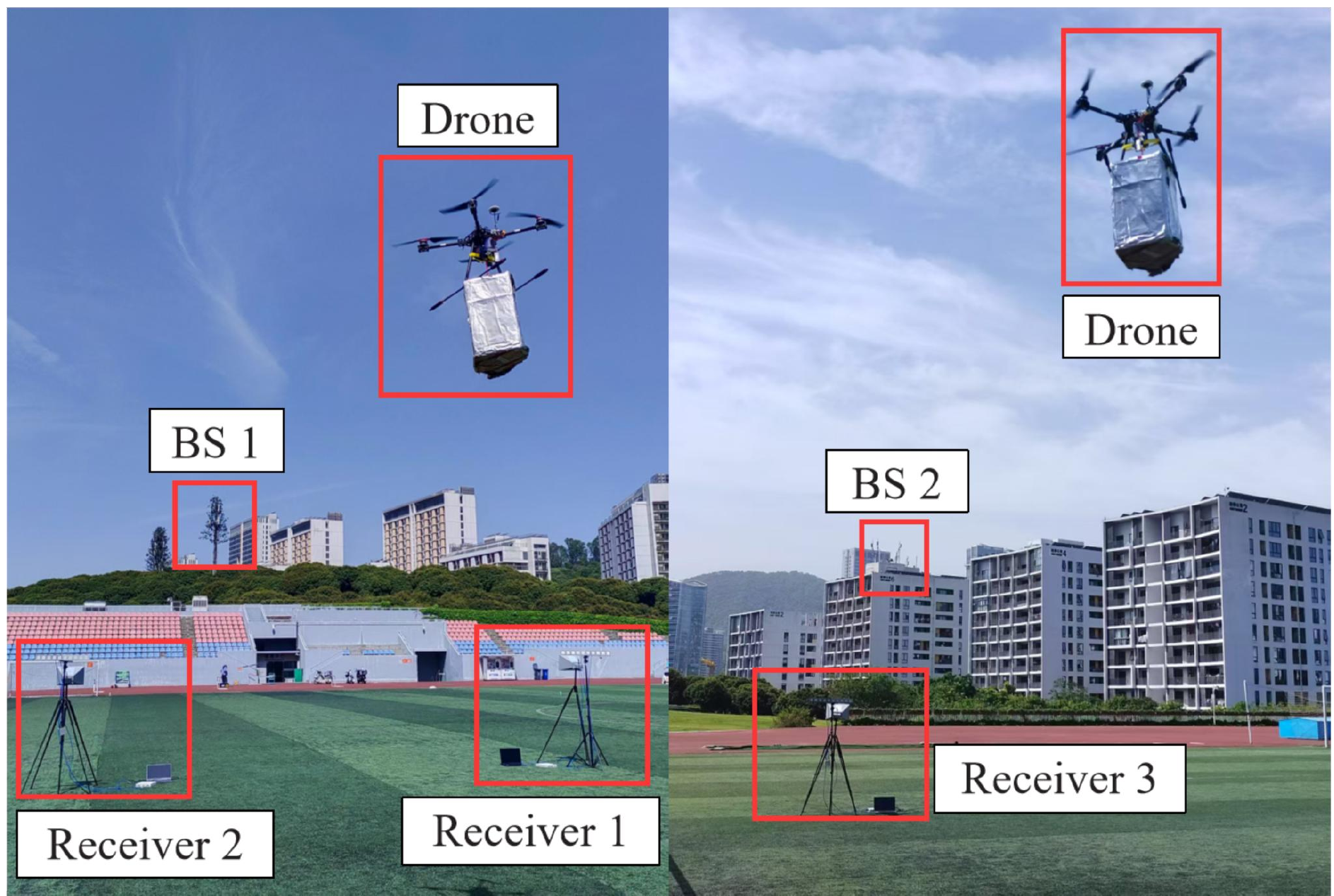}%
    \label{fig4-b}}
    \caption{Illustration of the experiment scenario.}
     \vspace{-0.5cm}
\end{figure}

\begin{figure}[!ht]  
    \centering
    \vspace{-0.1cm}
    \subfloat[Receiver 1]{
        \includegraphics[width=0.9\linewidth]{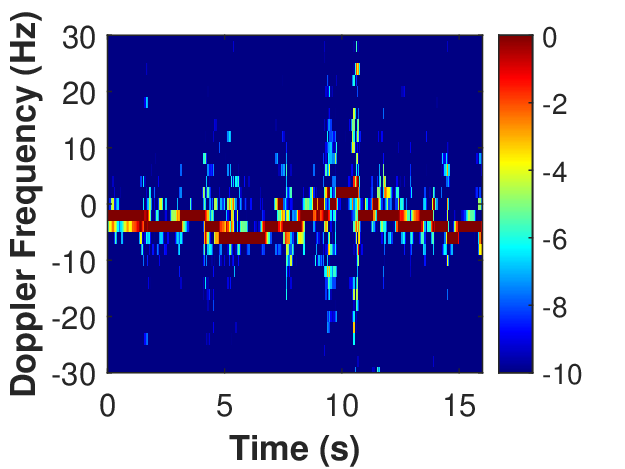}
    }\hfil \vspace{-0.03cm}
    \subfloat[Receiver 2]{
        \includegraphics[width=0.9\linewidth]{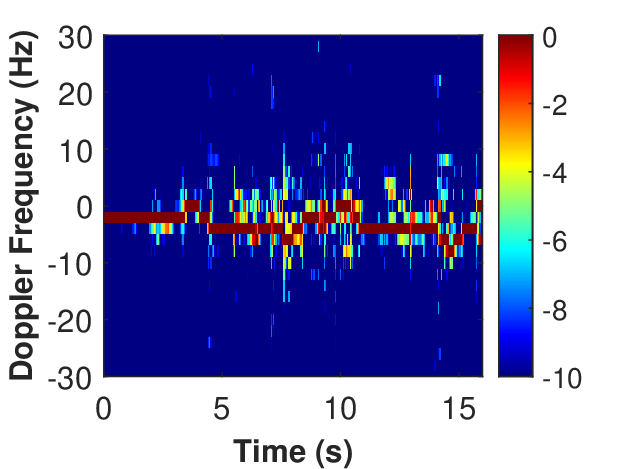}
    }\hfil \vspace{-0.03cm}
    \subfloat[Receiver 3]{
        \includegraphics[width=0.9\linewidth]{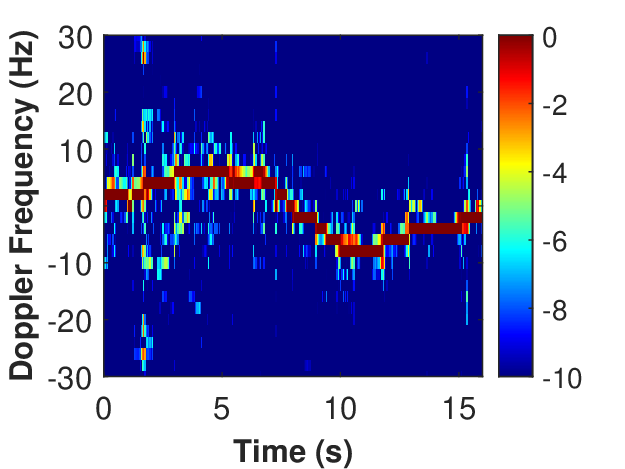}
    }
    \vspace{-3pt}
    \caption{The time-Doppler spectrograms at the three receivers.}
    \label{fig5}
    \vspace{-0.5cm}
\end{figure}

\section{Experiments and Discussions} \label{sec5}

\begin{figure}[!ht] 
    \centering
    \vspace{-0.1cm}
    \subfloat[V-shaped]{
        \includegraphics[width=0.9\linewidth]{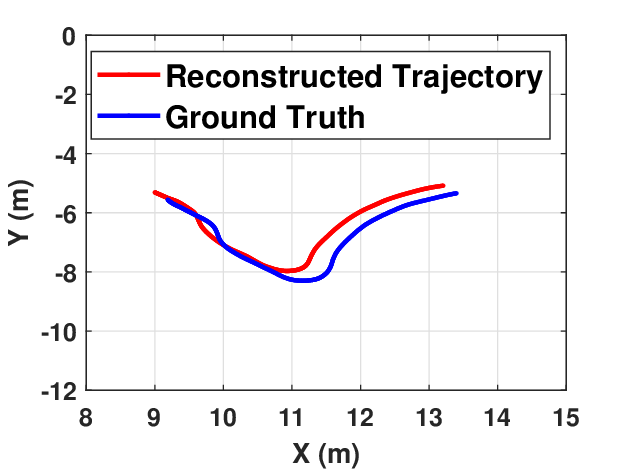}
    }\hfil \vspace{-0.03cm}
    \subfloat[L-shaped]{
        \includegraphics[width=0.9\linewidth]{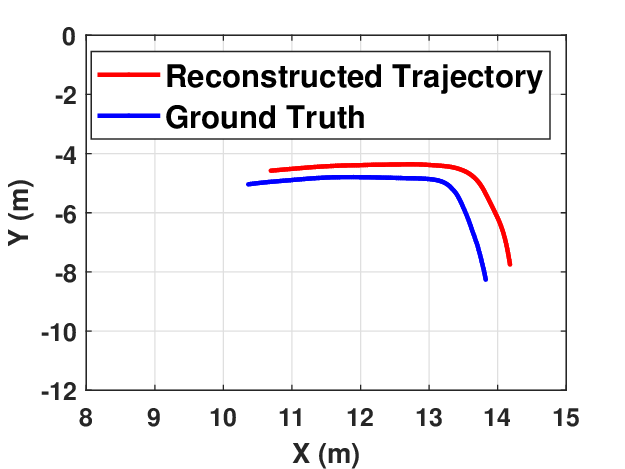}
    }\hfil \vspace{-0.03cm}
    \subfloat[U-shaped]{
    \includegraphics[width=0.9\linewidth]{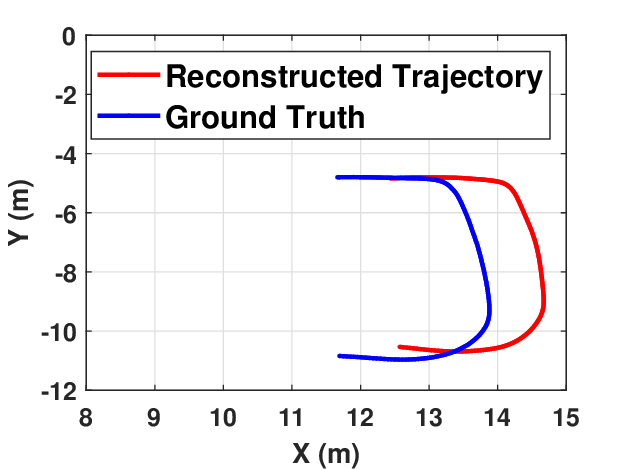}
    }
    \vspace{-3pt}
    \caption{Tracking results of the three trajectories, where the blue and red curves refer to the ground truths and the reconstructed trajectories without the knowledge of initial location.}
    \label{fig7}
    \vspace{-0.5cm}
\end{figure}
The performance of the proposed drone tracking system is demonstrated in a stadium, as shown in Fig. \ref{fig4-a}. Each receiver consists of a horn antenna, a linear array, a USRP-2953R and a laptop. The horn antenna and the linear array are used for the reference channel and the surveillance channel, respectively. The received signals of both channels at each receiver are sampled at the USRP and processed by the laptop. In the signal processing, $N_{w}\mathrm{T_{s}}=0.5$s and 
$\mathrm{T_{d}}=0.05$s. Thus, the resolution of the bistatic Doppler frequency is
$1/N_{w}\mathrm{T_{s}} =2$Hz, and the resolution of the bistatic range rate is around $0.16$m/s.

The experimental scenario is shown in Fig. \ref{fig4-b}. Three receivers are deployed on a soccer field, while two LTE BSs are at different locations: one is on a hill and the other is on the rooftop of a building. There are a number of buildings nearby, creating a number of static non-line-of-sight (NLoS) paths. The distances between the BSs and their corresponding receivers are around $190$m, $190$m and $230$m, respectively. The carrier frequencies of two LTE BSs are $1.85$GHz and $1.87$GHz, respectively. Both BSs work in Frequency-Division Duplex (FDD) mode with a downlink bandwidth of $20$MHz. However, a baseband signal with the bandwidth of $2.5$MHz is filtered from the $20$MHz downlink signal for the detection of bistatic Doppler frequency. Note that the proposed system tracks the drone's trajectory without range detection, a small signal bandwidth could save the hardware cost. 

A quadcopter suspended with a cardboard can be regarded as a delivery drone. In the experiments, three different flight trajectories are evaluated, namely V-shaped, L-shaped and U-shaped as illustrated in Fig. \ref{fig7}. In order to acquire the ground truths of the trajectory, the drone is equipped with an RTK positioning module with a centimeter-level localization accuracy. When solving the tracking problem in (\ref{eq8}), $495$, $405$ and $595$ initial locations for the three trajectories are chosen from the drone tracking area, respectively. For each initial location, five candidate trajectories are randomly initialized.

The CAFs at the three receivers versus time are illustrated in Fig. \ref{fig5}, where the target drone follows the V-shaped trajectory. The amplitudes of CAF are differentiated by colors. It can be observed that due to the different signal propagation paths of the three surveillance channels, the bistatic Doppler frequencies exhibit distinct patterns at the three receivers. The signs of the Doppler frequencies indicate the drone's moving direction. 

\begin{figure}[!ht]
    \centering
    \vspace{-0.4cm}
        \includegraphics[width=0.48\textwidth]{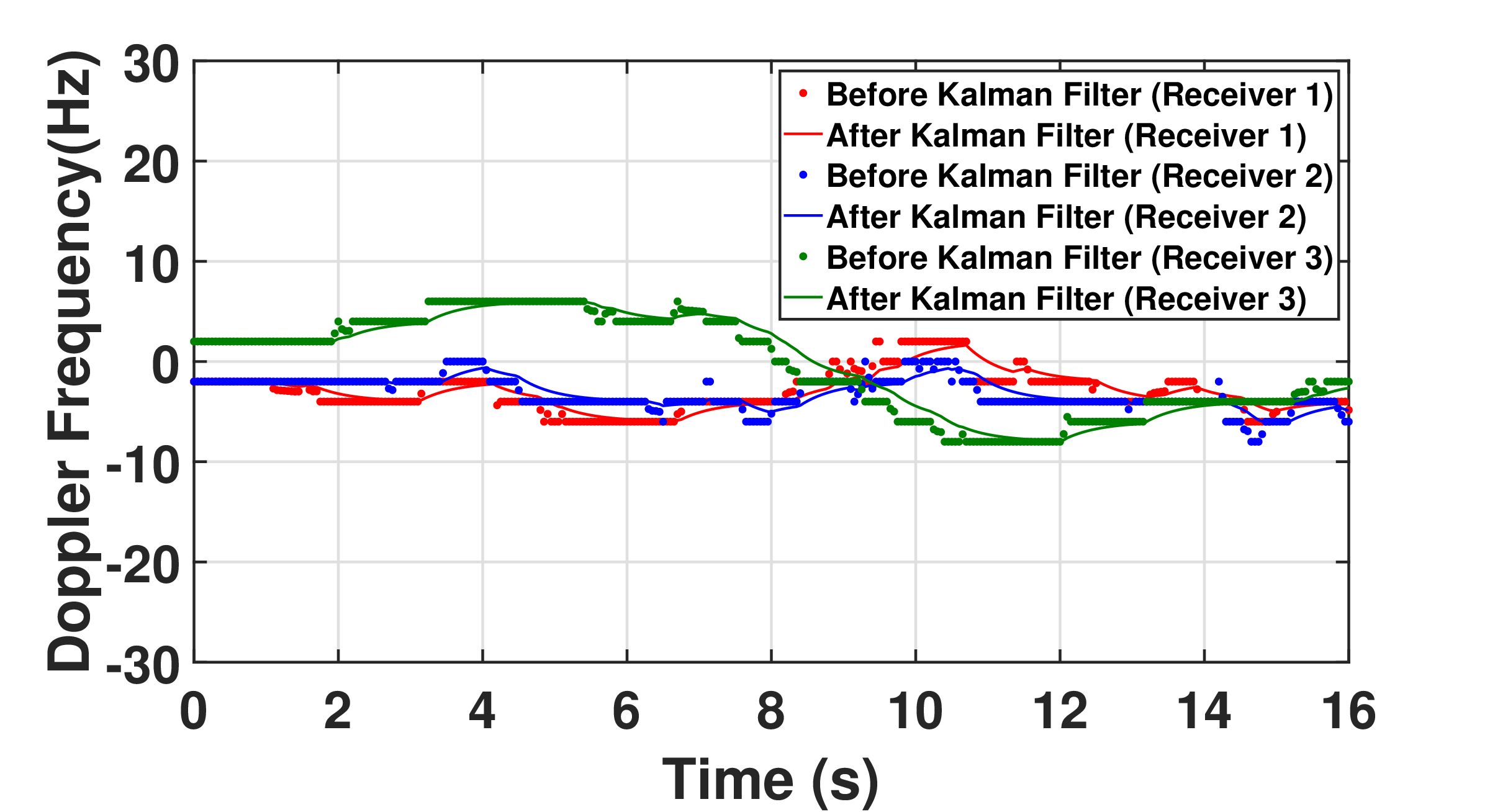}
    \vspace{-6mm}
    \caption{Detected and smoothed bistatic Doppler frequencies of the three receivers.}
    \label{fig6}
    \vspace{-0.7cm}
\end{figure}

Meanwhile, the estimations of bistatic Doppler frequency at the three receivers before and after Kalman filter are shown in Fig. \ref{fig6}. Compared to the time-Doppler spectrograms in Fig. \ref{fig5}, it can be observed that the false alarms and miss detections can be effectively suppressed after Kalman filter, providing reliable estimations for trajectory tracking.

\begin{figure}[!ht]
    \centering
    \includegraphics[width=0.47\textwidth]{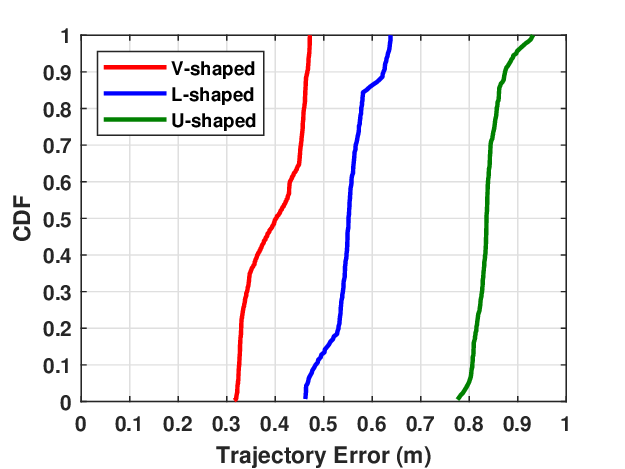}
    \vspace{-2mm}
    \caption{The CDFs of trajectory tracking error.}
    \label{fig8}
    \vspace{-0.3cm}
\end{figure}

The reconstructed trajectories and the ground truths
are compared in Fig. \ref{fig7}. It can be observed that the three reconstructed trajectories are generally consistent with the ground truths in shape, though some deviations occur due to estimation error. Moreover, the cumulative distribution functions (CDFs) of the tracking errors are shown in Fig. \ref{fig8}. It can be found that for all the above trajectories, $90\%$ of the tracking errors are less than $0.9$m.

\section{Conclusions} \label{sec6}
In this paper, a multistatic drone tracking  system via downlink LTE signals is proposed and demonstrated. The system consists of two LTE BSs and three receivers, each receiver detects the bistatic Doppler frequencies of the target drone at different directions. With the detected bistatic Doppler
frequencies of three different directions, the trajectories of the drone can be estimated by solving a minimum mean-squared problem. Thus, the angle and range detection in the existing drone tracking works are not necessary, reducing the implementation cost significantly. The experimental results demonstrate that when the target drone is around $200$m away from the LTE BSs and $20$m from the sensing receivers, $90\%$ of the tracking errors are less than $0.9$m. Without any signal emission, the proposed system can be densely deployed on urban rooftops to enable fine-grained drones monitoring.

\scriptsize
\bibliographystyle{IEEEtran}
\bibliography{main.bib}
\end{document}